\documentclass[a4paper]{IEEEtran}

\usepackage{graphicx}
\usepackage{pgfplots,amsmath,amssymb}
\usepackage{amsthm}
\usepackage{mathtools}

\newcommand{\codesmlc}{$\mathtt{sMLC}$}
\newcommand{\codeumlc}{$\mathtt{uMLC}$}
\newcommand{\codesbicm}{$\mathtt{sBICM}$}
\newcommand{\codeubicm}{$\mathtt{uBICM}$}

\newlength{\figwidth}
\newlength{\figheight}
\newlength{\lnwi}
\begin{document}

\title{Probabilistically Shaped Multi-Level Coding with Polar Codes for Fading Channels}

\author{\IEEEauthorblockN{Onurcan \.{I}\c{s}can, Ronald B\"ohnke, Wen Xu\\}
	\IEEEauthorblockA{Huawei Technologies D\"usseldorf GmbH, German Research Center\\
		Email: \{Onurcan.Iscan, Ronald.Boehnke, Wen.Dr.Xu\}@Huawei.com }}

\maketitle

\begin{abstract}
A probabilistic shaping method for multi-level coding (MLC) is presented, where the transmitted symbols are forced to have a shaped non-uniform distribution. 
It is shown that shaping only a single bit-level suffices to compensate for most of the shaping loss on the fading channels. A polar code based implementation of the proposed scheme is presented, where shaping is performed by using a precoder at the transmitter without increasing the decoding complexity. Simulation results show that performance improvements can be obtained compared to BICM- and MLC-based polar coding without shaping on Rayleigh fading channels.\footnote{This work is accepted for publication at Globecom 2018 Workshops. Copyright IEEE 2018.}
\end{abstract}


\section{Introduction}
In order to approach the theoretical capacity limits in a communication system, the transmitted symbols should in general  have a non-uniform distribution. However, many existing systems use uniformly distributed transmit symbols, which can lead to a shaping loss. Probabilistic shaping (PS) is a promising approach that can reduce this loss for higher order modulation. 
For additive white Gaussian noise (AWGN) channels, it is shown in \cite{bocherer2015bandwidth} that bit-interleaved coded modulation (BICM) with independent demapping combined with PS can compensate for the demapping and shaping losses, and approach the capacity. A scheme was also given in \cite{bocherer2015bandwidth}, where an additional distribution matcher/dematcher is used as a shaping encoder/decoder together with systematic error correction codes. In \cite{lnt_hwdu} this idea is extended to polar codes, where a precoder is used for systematic encoding, allowing the utilization of distribution matchers similar to \cite{bocherer2015bandwidth}. Moreover, a multi-level demapping and successive decoding approach is utilized to avoid demapping loss.

Recently, another PS approach with polar coding for higher order modulation is introduced in \cite{shaped_polar,icscan2018polar} (by extending the ideas from \cite{honda2013polar}), where a polar decoder is used as a precoder for signal shaping. The idea is to generate some shaping bits using the precoder and append them to the information bits prior to the polar encoding, such that the bits in the generated codewords have a certain probability distribution. Codewords with non-uniform distribution of bits are then mapped to channel input symbols, such that they have a desired distribution resulting in a shaping gain. Compared to PS approaches that use non-binary distribution matchers (such as \cite{bocherer2015bandwidth} and \cite{lnt_hwdu}), this approach cannot compensates for the full shaping loss, since only an approximation of the optimal symbol distribution can be obtained. However, it is shown in \cite{shaped_polar} that still a large shaping gain can be obtained. Moreover, this approach does not require a distribution dematcher at the receiver, as the shaping bits are treated as regular information bits at the receiver that can be discarded after decoding. 

The existing literature on PS mainly considers AWGN channel model (see \cite{bocherer2015bandwidth,wachsmann1999multilevel} and references therein). 
Here, we aim to develop a solution for higher order modulation with non-uniform distribution for transmission over fading channels. We first discuss different transmission strategies in Sec. \ref{sec:highmod} and show that BICM based PS (with independent demapping) cannot fully remove the shaping loss on fading channels (unlike on AWGN channels). We then propose in Sec. \ref{sec:sbs} a \textit{shaped multi-level coding} (s-MLC) approach with single bit-level shaping (SBS) that can remove most of the loss. Motivated by the asymptotic results, we build polar codes by extending ideas from \cite{honda2013polar} and \cite{shaped_polar} (in Sec. \ref{sec:polar_SBS}), where shaping gain is obtained by modifying mainly the transmitter. Our simulation results indicate that the proposed scheme can significantly outperform conventional BICM and MLC (with uniform distribution) based on the 5G New Radio (NR) polar codes \cite{chan_code5G}. 

In this work, we use lowercase bold letters (e.g. $\mathbf{x}$) for vectors and uppercase letters (e.g. $X$) for the random variables representing the elements of the associated vectors. $\text{P}_{X}(x)$, $\mathbb{H}(X)$ and $\mathbb{E}(X)$ denote the probability distribution, entropy and expected value of $X$. $\mathbb{I}(X;Y)$ is the mutual information between $X$ and $Y$, and 
BSC($p$) represents a binary symmetric channel with crossover probability $p$. Calligraphic letters (e.g. $\mathcal{F}$) denote sets.

\section{Transmission over Fading Channels with Higher Order Modulation}
\label{sec:highmod}
Consider the real-valued channel model 
$\mathbf{y} = \mathbf{h} \odot \mathbf{x} + \mathbf{w}$, 
where $\odot$ denotes element-wise multiplication. $\mathbf{h}$ and $\mathbf{w}$ contain the fading coefficients and the Gaussian noise samples, respectively. $\mathbf{x}$ contains the $2^m$ amplitude shift keying (ASK) channel input symbol taken from the alphabet 
$\mathcal{X}_m=\{\pm 1, \pm 3,\cdots,\pm(2^m-1)  \}$. The signal to noise ratio (SNR) is $\gamma = { \mathbb{E}( X^2 )  }/{ \mathbb{E}( W^2 )}$.
A symbol mapper generates $\mathbf{x}$ from $\mathbf{c}_i$, $i=1,\cdots,m$, containing the bits in the $i$th ASK bit-level. The extension to a complex-valued channel model is straightforward. 
The maximum achievable rate with perfect channel state information (CSI) at the receiver is 
given by $R = \mathbb{I}(X;Y|H)$. In the rest of the paper, we consider Rayleigh distributed fading coefficients that are known to the receiver, i.e., each ASK symbol is subject to a different independent fading coefficient. Note that $R$ is maximized with a Gaussian distributed $X$ for continuous input alphabets, and with Maxwell-Boltzmann (MB) distribution with $\text{P}_{X}(x) \sim  e^{-\nu x^2}$ for non-continuous input alphabets.  Below we consider different transmission strategies.

\begin{figure}[t]
	\centering
	\includegraphics{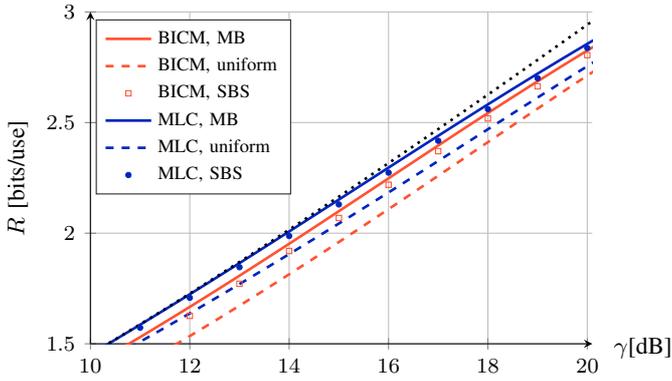}
	\caption{Achievable rates on Rayleigh fading channels for $m=4$ using MLC and BICM with uniform, MB and approximate MB (with single bit-level shaping) distributed ASK symbols. The parameters of the non-uniform distributions are numerically optimized at each SNR to maximize the achievable rate with the respective transmission scheme. Dotted line shows the channel capacity.}
	\label{fig:rates_m4}
\end{figure}

\subsection{Bit-Interleaved Coded Modulation}
$k$ information bits are encoded by a channel code to obtain a binary codeword $\mathbf{c}$ of length $n=m n_c$, which is interleaved and partitioned to  $\mathbf{c}_i$, $i=1,\cdots,m$, i.e., each ASK bit-level contains bits from the same codeword. The transmission rate is $R=k/n_c$, and with independent demapping, non-negative rates up to  
\begin{align}
\label{eq:BICMrate}
R_{\text{BICM}} = \mathbb{H}(X) - \textstyle\sum_{i=1}^{m}\mathbb{H}(C_i|Y,H)
\end{align}
can be achieved \cite{bocherer2017achievable}. Fig.  \ref{fig:rates_m4} shows $R_{\text{BICM}}$ with uniform and MB symbol distribution for $m=4$ with Gray labeling, which is typically used with BICM \cite{caire1998bit}.
We optimized parameters of the MB distribution numerically to maximize the achievable rate at each SNR. We observe that BICM with MB distributed channel input symbols outperforms uniform BICM, but there is still a gap to the capacity. Recall that BICM with probabilistic shaping can approach the capacity of AWGN channels, if the symbol distribution is optimized for the target SNR. However, it suffers from the independent demapping loss due to the variation of the instantaneous SNR in fading channels, i.e. signal shaping does not completely compensate for the loss, because the dependence between bit-levels are not taken into account during demapping.

\subsection{Multi-Level Coding}
$k$ information bits are first divided into $m$ parts with lengths $k_i$, $i=1,\cdots,m$, which are encoded to codewords $\mathbf{c}_i$ of length $n_c$ and transmitted over the $i$th ASK bit-level. 
The transmission rate is $R=\sum_{i=1}^{m}k_i/n_c$.
The receiver performs multi-level demapping and successive decoding, i.e., after decoding each bit level, this information is used for demapping the next bit-levels. The achievable rate is \cite{wachsmann1999multilevel}
\begin{align}
\label{eq:MLCrate}
R_{\text{MLC}} = \sum_{i=1}^{m} \underbrace{ \mathbb{I}(C_i;Y|C_1,\cdots,C_{i-1},H)}_{I_i}.
\end{align}
Observe that $R_{\text{MLC}}$ is the sum of bit-level capacities $I_i$, and the choice of the binary labeling does not affect the sum-rate, but only influences the rate of each individual bit-level. Therefore, asymptotically any binary labeling is optimal. 
Fig.~\ref{fig:rates_m4} shows $R_{\text{MLC}}$ with uniform and MB channel input symbol distribution for $m=4$, where the parameters of the MB distribution is optimized numerically for each SNR. We observe that MLC with MB distributed channel input symbols performs best, which motivates us to use signal shaping with MLC for fading channels.

Although the choice of the bit labeling does not influence the achievable rates asymptotically, it can have an effect on the performance in finite lengths. In \cite{wachsmann1999multilevel} and \cite{seidl2013polar} it is shown that a large variance of the bit-level capacities is beneficial in finite lengths, i.e., capacity of each bit-level should be preferably different. Therefore, we use set-partitioning labeling, which corresponds to natural binary code (NBC) for ASK, guaranteeing non-decreasing minimum Euclidean distance (and thus a non-decreasing capacity) for each bit-level if successive decoding is performed starting from the least significant bit. 

\section{Signal Shaping for Multi-Level Coding}
\label{sec:sbs}
In the presented schemes, the channel input symbols are obtained from binary codewords.
To obtain symbols with an MB distribution, the binary codewords transmitted on some bit-levels need to have non-uniform distribution of bits, i.e., $\text{P}_{C_i}(1)\neq 0.5$. Moreover, the bit-level distributions may need to be conditioned on each other, which implies joint encoding of bit-levels. 
By relaxing the condition on joint encoding, one can approximate the MB distribution by a product distribution (as in bit-level probabilistically shaped coded modulation, PSCM \cite{pikus2017bit}), where each bit-level is encoded independently with $\text{P}_{C_i}(1)=p_i$. 
In the following, we show that having only a single bit-level with non-uniform distribution suffices to generate an approximation of the MB distribution that can compensate for most of the shaping loss on fading channels. 
\subsection{Single Bit-Level Shaping (SBS)}
We consider the following piecewise constant approximation of the MB distribution for $2^m$-ASK symbols
\begin{align}
\label{eq:pmf}
\text{P}_{X}(x) = \left\{
\begin{array}{ll}
  p/2^{m-1}  &\text{if } |x|<2^{m-1}\\
	(1-p)/2^{m-1} & \text{otherwise.}
\end{array}
\right.
\end{align}
This distribution requires only a single bit-level $t$ with $\text{P}_{C_t}(1)=p\geq0.5$, where this bit-level determines whether $|x|$ is smaller than  $2^{m-1}$ or not, i.e., $c_t=1$ for $|x|<2^{m-1}$. 
Accordingly, symbols with high energy (i.e. with $|x|>2^{m-1}$) are transmitted less likely than the symbols with low energy.

\begin{figure}[t]
	\centering
	\includegraphics{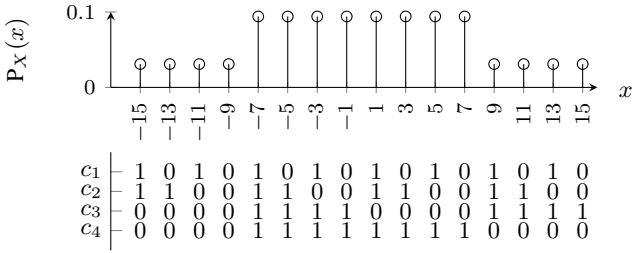}
	\caption{Shifted NBC labeling for $2^4$-ASK and resulting $\text{P}_X(x)$ for $p=0.75$.}
	\label{fig:pmf}
\end{figure}

A labeling with such a property can be obtained from NBC by a circular shift of $2^{m-2}$, as shown in Fig. \ref{fig:pmf} for $m=4$ and $t=4$. We call this labeling  \textit{shifted NBC}. The mapping from $x\in\mathcal{X}_m$ can be described by the binary representation of $\left[ (2^m - 1 - x)/2 + 2^{m-2}  \mod 2^m\right]$ with most significant bit corresponding to the $m$th bit-level, and $t=m$.
Note that the circular shift does not change the Euclidean distances of the bit-levels. There exist also other equivalent labelings with the same distance properties. We leave further optimizations of the bit-labeling
for non-uniform distribution and finite lengths as future work.

Observe that the $m$th bit-level determines the shape of the distribution and $\text{P}_{C_m}(1)\neq 0.5$, i.e., the $m$th bit level requires special channel codes that we discuss in Sec. \ref{sec:polar_SBS}.
Note that channel encoders usually generate codewords with $\text{P}_{C_i}(1)=0.5$, thus there are no restrictions for the first $m-1$ bit-levels.

\subsubsection*{Achievable Rates for SBS}
\begin{figure}[t]
	\centering
	\includegraphics{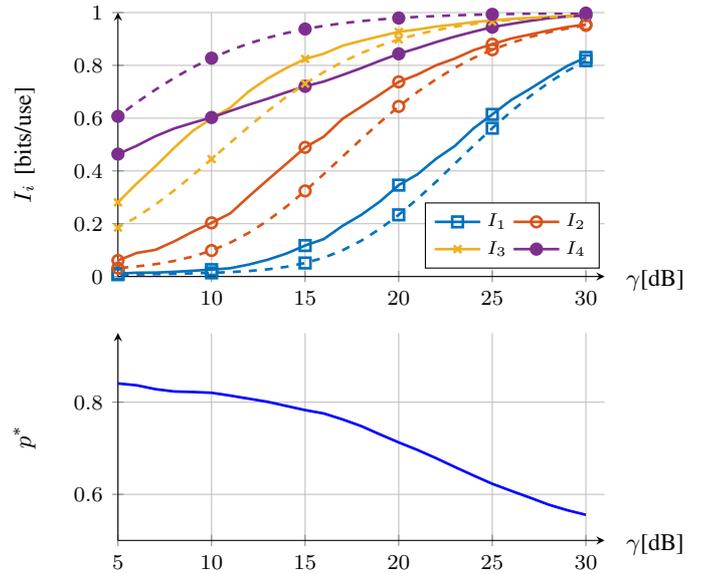}
	\caption{Numerically obtained $p^{*} = \arg\max_p R_{\text{MLC}}(\gamma)$ with the proposed bit-labeling (lower figure), and the bit-level capacities $I_i$ with the optimized $p^{*}$ (solid lines) and $p=0.5$ (dashed lines).}
	\label{fig:opt_p}
\end{figure}

For each operating SNR $\gamma$, we find by numerical search a $p^{*}$ that maximizes (\ref{eq:MLCrate}), shown in Fig. \ref{fig:opt_p} for $m=4$. 
Fig. \ref{fig:opt_p} also depicts the resulting bit-level capacities with optimized $p^{*}$ and with $p=0.5$. Note that the optimal $p^{*}$ leads to a decreased capacity of the last bit-level, but it also leads to increased capacities for other bit-levels, resulting in a larger sum-rate.
Fig. \ref{fig:rates_m4} shows the achievable rates of SBS-MLC, which are close to optimal. This implies that $R$ is not very sensitive with respect to the input distribution, and an approximation of the optimal distribution is nearly sufficient.

SBS can also be combined with Gray labeled BICM by using the method in \cite{icscan2018polar} and shaping only a single bit-level.
By maximizing (\ref{eq:BICMrate}) over $p$ we obtain the achievable rates for SBS-BICM, which are also given in Fig. \ref{fig:rates_m4} as reference.

\section{Polar Codes with Single Bit-Level Shaping}
\label{sec:polar_SBS}
Polar codes are known to work well with MLC \cite{seidl2013polar}, and they also allow generating codewords with $\text{P}_{C}(1)\neq 0.5$. In this work we use them in combination with  s-MLC.

\subsection{Polar Codes}
Channel polarization \cite{Arikan09} is an operation, where a physical channel is converted into virtual channels having either very high or very low reliabilities asymptotically. A polar encoder assigns information bits to reliable virtual channels, and (known) frozen bits to the unreliable ones. A  polar decoder (e.g. successive cancellation list (SCL) decoder \cite{Tal15}) processes a noisy observation of a codeword together with the frozen bits to estimate the information bits. 
A polar codeword $\mathbf{c}$ of length $n$ is obtained from a binary sequence $\mathbf{u}$ by 
\begin{align}
	\label{eq:polareq}
	\mathbf{c} = \mathbf{u}\mathbf{G},
\end{align}
where $\mathbf{G}$ is the $n\times n$ polar transform matrix. $\mathbf{u}$ contains information bits at the indices described by the set $\mathcal{I}$, representing the reliable virtual channels, and frozen bits at the indices described by $\mathcal{F}$. We will call $(n,\mathcal{I},\mathcal{F})$ a polar code of length $n$.
The reliabilities of the virtual channels can be determined using density evolution (DE) \cite{mori2009performance}.

\subsection{Polar Codewords with Non-Uniform Distribution of Bits}
Assuming independent and uniform choices of information and frozen bits, the numbers of ones and zeros in a codeword are roughly equal, i.e., $\text{P}_{C}(1)=0.5$. In \cite{honda2013polar} and \cite{mondelli2014achieve} a method was presented to obtain polar codewords with a target probability $\text{P}_{C}(1)\neq 0.5$ for transmission over asymmetric channels. The idea is to use some of the virtual channels (denoted by the set $\mathcal{S}$) to transmit \textit{shaping bits}, which are generated from the information and frozen bits. The shaping bits do not convey any new information, but they force the codeword to have a target probability $\text{P}_{C}(1)$. 

\begin{figure}[t]
	\centering
	\includegraphics{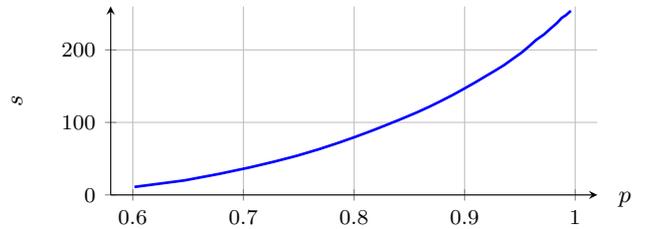}
	\caption{The required number of shaping bits $s$ to obtain codewords of length $n=256$ with $\text{P}_{C}(1)=p$, when an SCL decoder with list size 8 is used.}
	\label{fig:shapperf}
\end{figure}
\subsubsection*{Generation of Shaping Bits}
Assume $\mathbf{u}$ includes information, frozen and shaping bits at the indices $\mathcal{I}$, $\mathcal{F}$ and $\mathcal{S}$, respectively, and the resulting codeword $\mathbf{c}$ after the polar transform has $\text{P}_{C}(1)=p$. 
By reformulating (\ref{eq:polareq}) as $\mathbf{u}\mathbf{G} \oplus \mathbf{c} = \mathbf{0}$, we can interpret this as a polar codeword $\mathbf{u}\mathbf{G}$ being transmitted over a BSC($p$), where $\mathbf{c}$ is the noise vector and the channel output is the all-zero vector. 
Accordingly, if we utilize a polar decoder by treating the all-zero vector as the noisy observation and the bits at $\mathcal{I}\cup\mathcal{F}$ as frozen bits, we can obtain the shaping bits at the output of the decoder if $(n,\mathcal{S},\mathcal{I}\cup\mathcal{F})$ forms a polar code for a BSC($p$), as described in \cite{shaped_polar}.
\subsubsection*{Example 1} Consider transmitting $k$ information bits over a target channel with $n\rightarrow \infty$ and aim to have $\text{P}_{C}(1)=p$.
Let the elements of $\bar{\mathbf{e}}$ represent the reliabilities of the virtual channels if a BSC($p$) is polarized (obtained e.g. using DE). Similarly, let the elements of $\mathbf{e}$ denote the reliabilities of the target channel after polarization.
Note that asymptotically $\mathcal{S}$ contains the most reliable $s=\lfloor n(1-\text{h}_2(p))\rfloor$ virtual channels according to  $\bar{\mathbf{e}}$ \cite{shaped_polar}, where $\text{h}_2(p)$ is the binary entropy function.
$\mathcal{I}$ is constructed from the $k$ most reliable virtual channels of $\mathbf{e}$ that are in the set $\mathcal{N} \backslash \mathcal{S}$. The remaining virtual channels form $\mathcal{F}$.
For simplicity, one can assume $\bar{\mathbf{e}}=\mathbf{e}$, as for many binary symmetric channels the order of the reliabilities of the virtual channels is similar. As a result, if an ordered set of virtual channels is available (such as defined in \cite{chan_code5G}), one can use the most reliable $s$ channels for shaping bits, the least reliable channels for frozen bits and the rest for information bits.

Let $\mathbf{d}$ and $\mathbf{f}$ represent the vectors containing the information and frozen bits, respectively.
We use a polar decoder as a precoder to obtain the shaping bits $\mathbf{s}$, where $\mathbf{d}$ and $\mathbf{f}$ are treated as frozen bits, and $L_s=\log((1-p)/p)$ is used as the $L$-values representing the all-zero sequence. The decoder output would contain the shaping bits $\mathbf{s}$. Note that $\mathbf{d}$, $\mathbf{f}$ and $\mathbf{s}$ form $\mathbf{u}$, which can be fed to the polar transform as in (\ref{eq:polareq}) to generate the codeword $\mathbf{c}$ with $\text{P}_{C}(1)=p$. At the receiver, $\mathbf{s}$ is treated as information bits (like $\mathbf{d}$) that are discarded after decoding.  

\subsubsection*{Example 2} For finite lengths, the assumption about the number of the shaping bits may be inaccurate. Therefore, we evaluate $\text{P}_{C}(1)$ numerically. For $n=256$, we choose a target $p$, select an initial value for $s$ and randomly generate $n-s$ bits  that are used as information and frozen bits. 
Then we use the most reliable $s$ virtual channels from the ordered set in \cite{chan_code5G} as $\mathcal{S}$, and use an SCL decoder with list size 8 as a precoder to obtain $\mathbf{s}$, where we treat all-zero vector as the output of a BSC($p$). We use a polar transform to generate the codeword from the information, frozen and shaping bits, and evaluate $\text{P}_{C}(1)$ by Monte-Carlo simulations. We seek for $s$ that minimizes $|\text{P}_{C}(1)-p|$, which is shown in Fig. \ref{fig:shapperf}. We observe that with increasing number of shaping bits, the distribution of ones and zeros within the codeword becomes more non-uniform. 

\subsection{Shaped Multi-Level Coding with Polar Codes}
We propose using s-MLC with polar codes, where we employ conventional polar codes for the first $m-1$ bit levels and a polar code with $\text{P}_{C}(1)=p$ for the $m$th bit-level, all having codeword length $n_c$.
To improve the finite length performance, we use $z_i$ cyclic redundancy check (CRC) bits at each bit-level and use CRC-aided decoding \cite{Tal15}. The shaping bits at the $m$th bit-level are generated after CRC bits are appended.

The design of polar codes requires obtaining the reliabilities of the polarized virtual channels, which depends on the operating SNR. Similarly, the set of virtual channels used for the transmission of the shaping bits depends on the target distribution. To simplify the design process, we use the ordered set of virtual channels from \cite{chan_code5G}, which is specified for 5G NR control channels. Accordingly, for the bit-levels $i=1,\cdots, m-1$, the most reliable $k_i+z_i$ virtual channels are used for transmission of information and CRC bits and the rest is used for transmission of the frozen bits. For the $m$th bit-level, the most reliable $s$ virtual channels are used to transmit the shaping bits. From the remaining virtual channels, the most reliable ones are used for transmitting $k_m+z_m$ information and CRC bits, leaving the rest for frozen bits.

Note that s-MLC with polar codes requires only small modifications compared to a conventional MLC. The receiver takes $\text{P}_{X}(x)$ into account during demapping, and the decoded shaping bits are discarded at the receiver. Thus, the complexity increase at the receiver is negligible. The transmitter, however,  needs to run an additional polar decoder to generate the shaping bits. 

Below we discuss the choice of the parameters such that the number of transmitted information bits is maximized at a given operating SNR $\bar{\gamma}$, a target block error rate (BLER) $P_e$ and $n_c$ channel uses.

\subsubsection*{Choice of $p$}
The optimal $p^{*}$ can be obtained numerically by maximizing $R_{\text{MLC}}$ at $\bar{\gamma}$, as given in  Fig. \ref{fig:opt_p} for $m=4$. 

\subsubsection*{Choice of $s$}
After $p^{*}$ is determined, we use the approach in Example 2 to obtain the required number of shaping bits $s$. 

\subsubsection*{Choice of $k_i$}
One can approximate $k_i$ as $\lfloor I_i n_c \rfloor$ for very large $n_c$. However, for finite lengths and with a target $P_e$, the approximation will be inaccurate. Let $P_i$ denote the probability of incorrect decoding of the $i$th bit-level, which can be obtained by Monte-Carlo simulations. The BLER can be written as $P_e = 1-\prod_{i=1}^m{(1-P_i)}$. We seek for $k_i$ that produces a smaller error probability than $P_i$, assuming each bit-level has the same error probability $P_i=1-\sqrt[m]{1-P_e}$. 

Let the elements of $\mathbf{e}_i$ denote the reliability of the virtual channels after polarization at the $i$th ASK bit-level, assuming the previous bit-levels are error-free decoded and demapped at $\bar{\gamma}$. For given decoding parameters (e.g. list size) we find the maximum value of $k_i$ such that the block error probability (obtained by Monte-Carlo simulations) is less than $P_i$, if the most reliable $k_i+z_i$ virtual channels (excluding the ones used to transmit the shaping bits in the $m$th level) are used to transmit information and CRC bits. 

\section{Numerical Evaluations and Discussions}
\label{sec:eval}
We designed an s-MLC scheme \codesmlc\, for $2^4$-ASK and shifted NBC labeling, as described in the previous section. We chose the operating SNR $\bar{\gamma}$ such that a target error probability $P_e\leq10^{-3}$ is obtained for $k=512$ information bits with $n_c=256$ channel uses. An SCL decoder with list size 8 was employed as the precoder and decoder, and pseudo-random sequences as frozen bits. We used $z_i$ CRC bits for the $i$th bit-level.  

To compare the performance of the proposed scheme, we use three reference transmission schemes with similar distribution of $X$. 
As a first reference, an MLC scheme \codeumlc\, with NBC labeling and uniform $\text{P}_X(x)$ was designed, where we used the same procedure as before except for the choice of $p^{*}=0.5$ and $s=0$. 
A second reference is the BICM scheme \codeubicm\, with Gray labeling and uniform $\text{P}_X(x)$, where we used codewords of length $mn_c$ and a different pseudo-random interleaver between encoder and symbol mapper for each codeword.
As a further reference, we designed a BICM scheme \codesbicm\, with Gray labeling and non-uniform $\text{P}_X(x)$ according to \cite{shaped_polar}, where $s$ bits were used for shaping a single bit-level, resulting in a distribution given in (\ref{eq:pmf}). 
Note that all four schemes have $R=2$ bits/use, and $z=16$ additional CRC bits in total are employed for CRC-aided decoding. 
The parameters are given in Table~\ref{tab:param}.
\begin{table}[t]
	\caption{Parameters of the Evaluated Codes}
	\label{tab:param}
	\vspace*{-0.8cm}
\begin{center}
	\resizebox{\columnwidth}{!}{%
	\begin{tabular}{c|c c c c c c c  c}
						 		& $m$ 	& $n_c$ 		&$p^{*}$	& $s$ 	& $k$ 	&  & $z$ &  \\
					\hline	
					\codeumlc& $4$ 	& $256$ 	   &	& 	& $512$ 	& \begin{tabular}{c}
						$k_1 = \,\,14$\\
						$k_2 = \,\,92$\\
						$k_3 = 185$\\
						$k_4 = 221$\\
					\end{tabular}  & $16$ & \begin{tabular}{c}
					$z_1=4$\\
					$z_2 = 4$\\
					$z_3=4$\\
					$z_4 = 4$\\
				\end{tabular} \\	
					 \hline	
					 \codesmlc& $4$ 	& $256$ 			&$0.75$	& $56$ 	& $512$ 	& \begin{tabular}{c}
						$k_1 = \,\,24$\\
						$k_2 = 112$\\
						$k_3 = 197$\\
						$k_4 = 179$\\
					\end{tabular} & $16$ & \begin{tabular}{c}
						$z_1=4$\\
						$z_2 = 4$\\
						$z_3=4$\\
						$z_4 = 4$\\
					\end{tabular} \\	
					\hline	
					\codeubicm& $4$ 	& $256$ 	           &       &       & $512$ 	&    			   & $16$ & \\
					\hline					
					\codesbicm& $4$ 	& $256$ 	           &  $0.78$     & $72$      & $512$ 	&    			   & $16$ & \\					

	\end{tabular}
}
\end{center}
\end{table}

\begin{figure}
	\centering
	\includegraphics{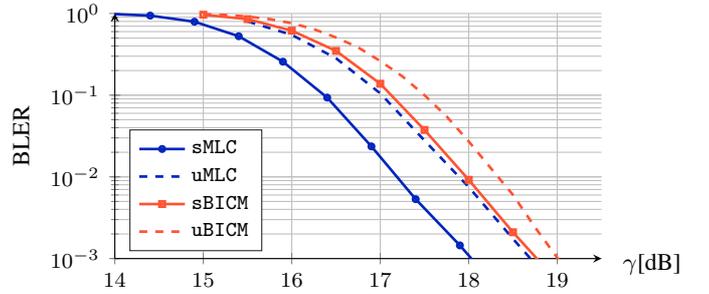}
	\caption{BLER performance on Rayleigh fading channels.}
	\label{fig:polar_bler}
\end{figure}

Fig. \ref{fig:polar_bler} shows the BLER performances. We observe that by using SBS instead of uniform signaling, the performance can be improved both for BICM and MLC. \codesmlc\, performs $0.7$dB better than \codeumlc\, and $1$dB better than \codeubicm, while the asymptotic gains (according to Fig. \ref{fig:rates_m4}) are $0.6$dB and $1.18$dB, respectively. Also note that MLC outperforms BICM, although shorter codewords (that cause a larger finite length loss) are employed. The proposed s-MLC with polar codes significantly outperforms the reference schemes.

Recall that for all designed codes we used the ordered set of virtual channels from \cite{chan_code5G} which is not necessarily optimal for any of the presented codes at the operating SNRs, and by evaluating the reliabilities of the virtual channels by using DE, one can improve the performance of each code. On the other hand, the performance curves with this suboptimal choice of virtual channels and the asymptotic results show similar trends.

In order to make a fair comparison, we only considered transmission strategies, where $X$ is distributed uniformly or according to (\ref{eq:pmf}). To obtain the full shaping gain on fading channels, one can use other transmission schemes resulting in an exact MB distribution with optimal parameters. For example, \cite{steiner2017ultra} proposes using non-binary channel codes together with non-binary distribution matchers and dematchers. This scheme would not suffer from a demapping loss on fading channels, but requires non-binary channel decoders, which are usually more complex than binary decoders. Another alternative is using the scheme presented in \cite{lnt_hwdu}, where an additional precoder (for allowing systematic encoding of polar codes by dynamic freezing) and non-binary distribution matchers and dematchers are required. 
Compared to \cite{lnt_hwdu} and \cite{steiner2017ultra}, the proposed scheme does not generate symbols with capacity achieving distribution, but as given in Fig. \ref{fig:rates_m4}, the expected additional gain by using an exact MB distribution is small. Moreover, the proposed scheme has a simple receiver structure, as it only requires small modifications to a conventional polar MLC receiver, and does not require a distribution dematcher. Moreover, the precoder at the transmitter can be realized by a polar decoder, which is already included in the transmission chain of bidirectional communication systems.  

\section*{Acknowledgment}
This work has been partly performed in the framework of the H2020 project ONE5G (ICT-760809) receiving funds from the EU. The views expressed in this work are those of the authors and do not necessarily represent the project view.

\bibliographystyle{IEEEtran}
\bibliography{IEEEabrv,mybibfile}

\end{document}